%% file: N_Erasmus.tex
\documentclass[twocolumn]{aastex62}
\usepackage{gensymb}
\usepackage[colorinlistoftodos]{todonotes}
\usepackage[normalem]{ulem}

\shorttitle{Investigating Taxonomic Diversity within Asteroid Families}
\shortauthors{Erasmus et al.}

\begin{document}

\title{\textbf{Investigating Taxonomic Diversity within Asteroid Families through ATLAS Dual-Band Photometry}}

\correspondingauthor{Nicolas Erasmus}
\email{nerasmus@saao.ac.za}

\author{N. Erasmus}
\affil{South African Astronomical Observatory, Cape Town, 7925, South Africa.}

\author{S. Navarro-Meza}
\affil{Instituto de Astronomia, Universidad Nacional Autonoma de Mexico, Ensenada B.C. 22860, Mexico.}
\affil{Department of Physics and Astronomy, Northern Arizona University, Flagstaff, AZ 86001, USA.}

\author{A. McNeill}
\affil{Department of Physics and Astronomy, Northern Arizona University, Flagstaff, AZ 86001, USA.}

\author{D. E. Trilling}
\affil{Department of Physics and Astronomy, Northern Arizona University, Flagstaff, AZ 86001, USA.}
\affil{South African Astronomical Observatory, Cape Town, 7925, South Africa.}

\author{A. A. Sickafoose}
\affil{South African Astronomical Observatory, Cape Town, 7925, South Africa.}
\affil{Department of Earth, Atmospheric, and Planetary Sciences, Massachusetts Institute of Technology, Cambridge, MA 02139-4307, USA.}
\affil{Planetary Science Institute, Tucson, AZ 85719-2395, USA}

\author{L. Denneau}
\affil{Institute for Astronomy, University of Hawaii, Honolulu, HI 9682, USA.}

\author{H. Flewelling}
\affil{Institute for Astronomy, University of Hawaii, Honolulu, HI 9682, USA.}

\author{A. Heinze}
\affil{Institute for Astronomy, University of Hawaii, Honolulu, HI 9682, USA.}

\author{J. L. Tonry}
\affil{Institute for Astronomy, University of Hawaii, Honolulu, HI 9682, USA.}

\begin{abstract}
We present here the \textit{c-o} colors for identified Flora, Vesta, Nysa-Polana, Themis, and Koronis family members within the historic data set (2015-2018) of the Asteroid Terrestrial-impact Last Alert System (ATLAS). The Themis and Koronis families are known to be relatively pure C- and S-type Bus-DeMeo taxonomic families, respectively, and the extracted color data from the ATLAS broadband \textit{c}- and \textit{o}-filters of these two families is used to demonstrate that the ATLAS \textit{c-o} color is a sufficient parameter to distinguish between the C- and S-type taxonomies. The Vesta and Nysa-Polana families are known to display a mixture of taxonomies possibly due to Vesta's differentiated parent body origin and Nysa-Polana actually consisting of two nested families with differing taxonomies. Our data show that the Flora family also displays a large degree of taxonomic mixing and the data reveal a substantial H-magnitude dependence on color. We propose and exclude several interpretations for the observed taxonomic mix. Additionally, we extract  rotation periods of all of the targets reported here and find good agreement with targets that have previously reported periods. 

\end{abstract}


\section{Introduction}
\label{sec:intro}

Asteroid families are groups of objects where members of the group display similar proper orbital elements. This suggests a common collisionally-disrupted parent-body source for a family.  Since the identification of asteroid families\citep{Hirayama1918}, several studies have also shown that for many of the identified families there is also a strong correlation between taxonomic type and family membership \citep{Bus1999}. For instance, family members of the Massalia, Eunomia, and Koronis families all have observed spectra or colors that are consistent with the Bus-DeMeo S-type taxonomy \citep{Erasmus2019,Masiero2015,Lazzaro1999}. On the other hand, members of the Hygiea,  Adeona, and Themis families all have observed spectra or colors that are consistent with the Bus-DeMeo C-type taxonomy \citep{Erasmus2019,Masiero2015, Carruba2013}. This leads to the assumption that the individual parent-body sources of each of these families must have had a pure composition with either a S-type or C-type spectral signature (see Figure 1 for spectra in the visible of the Bus-DeMeo S- and C-type taxonomies).  

However, there are several families that display a large variation in observed spectra or color and therefore potentially contain a significant mix of two (or more) taxonomies.  Some explanations that have been proposed include: nested families, i.e., two overlapping families in orbital space that are actually two separate families with differing taxonomy \citep{Cellino2001}; two colliding parents that had completely different mineralogy; size-dependent space-weathering that modifies the spectral shape of a subgroup within the family \citep{Brunetto2005}; a large differentiated parent body having variations in mineralogy \citep{Binzel1993}; or simply incorrectly assigning membership (sometimes referred to as interlopers).

In this work we present the \textit{c-o} colors and rotation periods of 1613 (414 Vesta, 494 Flora, 304 Nysa-Polana, 204 Themis, and 197 Koronis) main-belt targets. In Sections \ref{sec:obs} and \ref{sec:family} we briefly summarize the Asteroid Terrestrial-impact Last Alert System (ATLAS) system and the approach for identifying the relevant family members within its data set. In Section \ref{sec:rot_color} we explain how the colors are derived from the data set's dual-band photometry. The method for extracting color relies on the determination of the rotation periods, hence the latter are a convenient by-product of our analysis.  In Section \ref{sec:tax_det}  we discuss how we assign taxonomies to the targets. Section \ref{sec:validation} contains results of validation tests using previously reported taxonomies and rotation periods. We also show validation results using the ATLAS \textit{c-o} colors of the Themis and Koronis targets which have known and relatively pure taxonomic distributions.  In Sections \ref{sec:vesta_nysa} and \ref{sec:flora} we perform a more in-depth study on the observed taxonomic diversity of the Vesta, Nysa-Polana and Flora families and use this to draw conclusions on the origin of the Flora family by comparing to the Vesta and Nysa-Polana families which have known reasons for the observed mixture in taxonomies present.

\section{ATLAS Data}
\label{sec:obs}
Observations were performed between 2015 and 2018 by the Asteroid Terrestrial-impact Last Alert System (ATLAS)\footnote[1]{\url{http://atlas.fallingstar.com}}. Currently consisting of two units both located in Hawaii, ATLAS is designed to achieve a high survey speed per unit cost \citep{Tonry2018}. Its main purpose is to discover asteroids with imminent impacts with Earth that are either regionally or globally threatening in nature. To fulfill this, the two current ATLAS units scan the complete visible northern sky every night enabling it to make numerous discoveries in multiple astronomical disciplines, such as supernovae candidates discovery \citep{Prentice2018}, gamma ray burst phenomena \citep{Stalder2017}, variable stars \citep{Heinze_2018}, and asteroid discovery. Since its inception ATLAS has found 39 Potentially Hazardous Asteroids among the 370 near-Earth asteroids that it has discovered. All detected  asteroid astrometry and photometry are posted to the Minor Planet Center, while the supernova candidates are publicly reported to the International Astronomical Union Transient Name Server. 

\begin{figure}
	\begin{center}
		\includegraphics[width=0.5\textwidth]{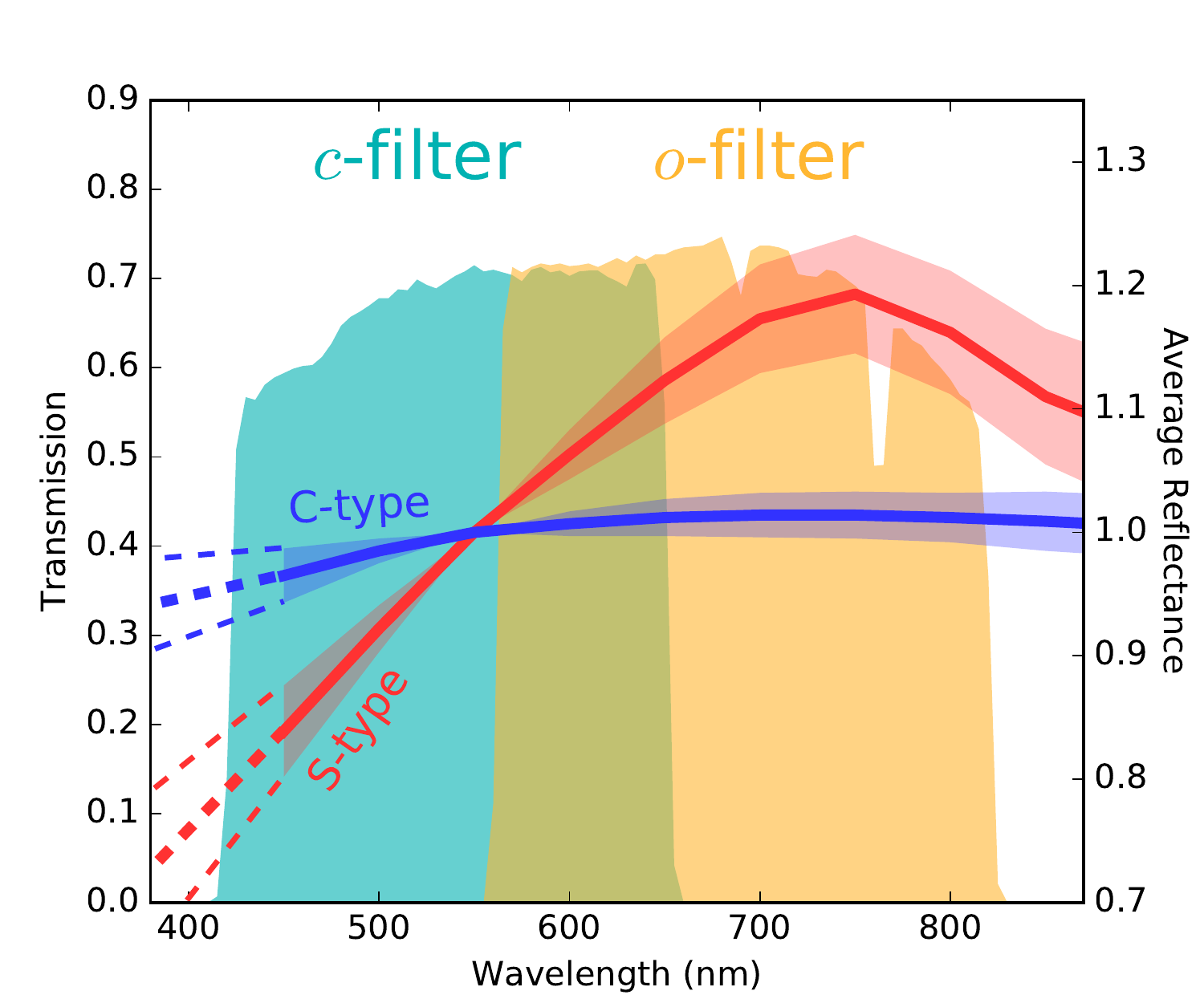}
		\caption{The transmission curves of the \textit{c}- and \textit{o}-filters of ATLAS \citep{Tonry2018} are plotted together with the averaged visible wavelength reflectance spectra, normalized at 550\,nm, of the Bus-DeMeo S- and C-type  taxonomies \citep{DeMeo2009}. The upper and lower bounds  of the two taxonomic spectra provided by \cite{DeMeo2009} are also indicated with shading.  The dashed lines are linear interpolations of the spectra and bounds since no data below 450\,nm is provided.}
		\label{filter_spectra}
	\end{center}
\end{figure}

The two ATLAS units are 0.5\,m telescopes each covering 30 deg$^2$ field-of-view in a single exposure. The main survey mode mostly utilizes two custom filters, a ``cyan'' or \textit{c}-filter with a bandpass between 420-650\,nm and an ``orange'' or \textit{o}-filter with a bandpass between 560-820\,nm (see Figure \ref{filter_spectra}). For further details on ATLAS, ATLAS photometry, and the ATLAS All-Sky Stellar Reference Catalog see \cite{Tonry2018,Heinze_2018,Tonry_2018b}. 

\section{Family Determination}
\label{sec:family}

To identify Flora, Vesta, Nysa-Polana, Themis, and Koronis family members within the ATLAS data set we utilize and cross-correlate with data from \cite{Nesvorny2015b} obtained through The Planetary Data System\footnote{\url{https://pds.nasa.gov/}} (PDS) to associate objects from the ATLAS data set with known collisional families. The \cite{Nesvorny2015b} data set makes use of the Hierarchical Clustering Method \citep{Zappala1990} to assign families. The data set also supplies a ``c-parameter'' (see \cite{Nesvorny2015b} for description) that can be used to identify suspected interlopers. For the ATLAS data set the interloper contamination is low with a percentage of suspected interlopes for the Flora, Vesta, Nysa-Polana, Themis, and Koronis of $4.5\%$, $0.0\%$, $13.5\%$, $6.4\%$ and $5.6\%$ respectively.

\section{Rotation Period Extraction and Color Calculation}
\label{sec:rot_color}

\begin{figure*}
	\begin{center}
		\gridline{ 
			\fig{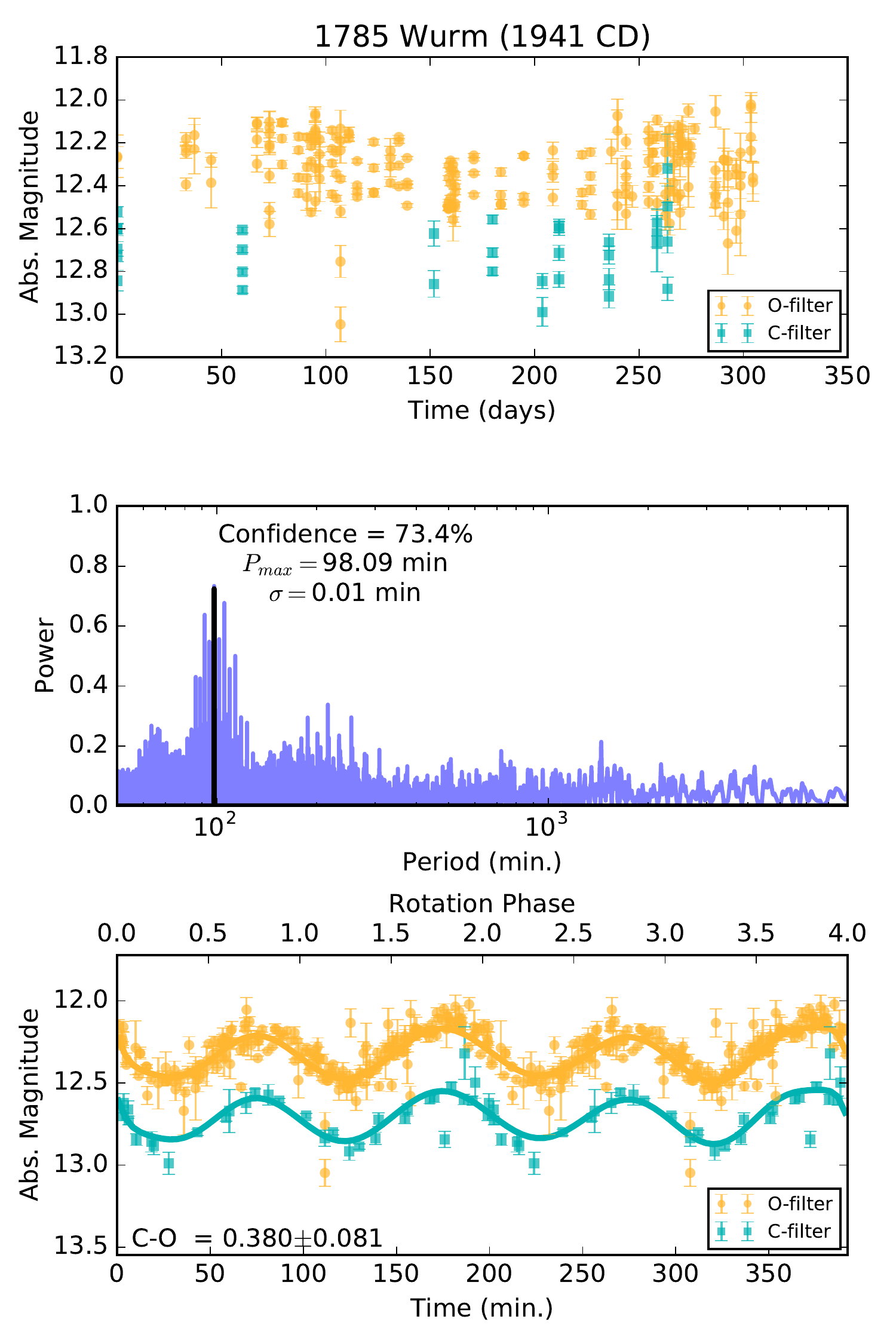}{0.3\textwidth}{(a)}
			\fig{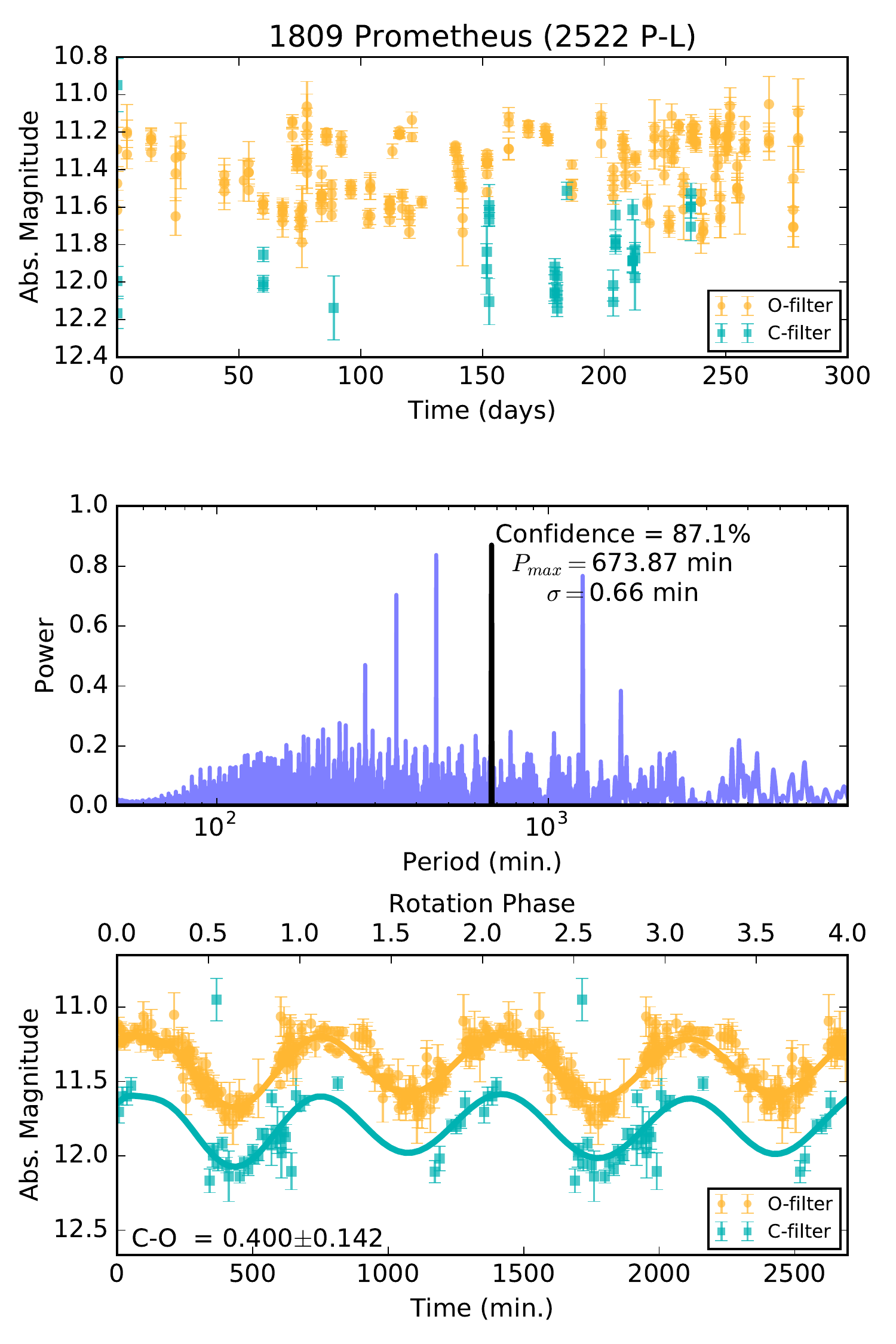}{0.3\textwidth}{(b)}
			\fig{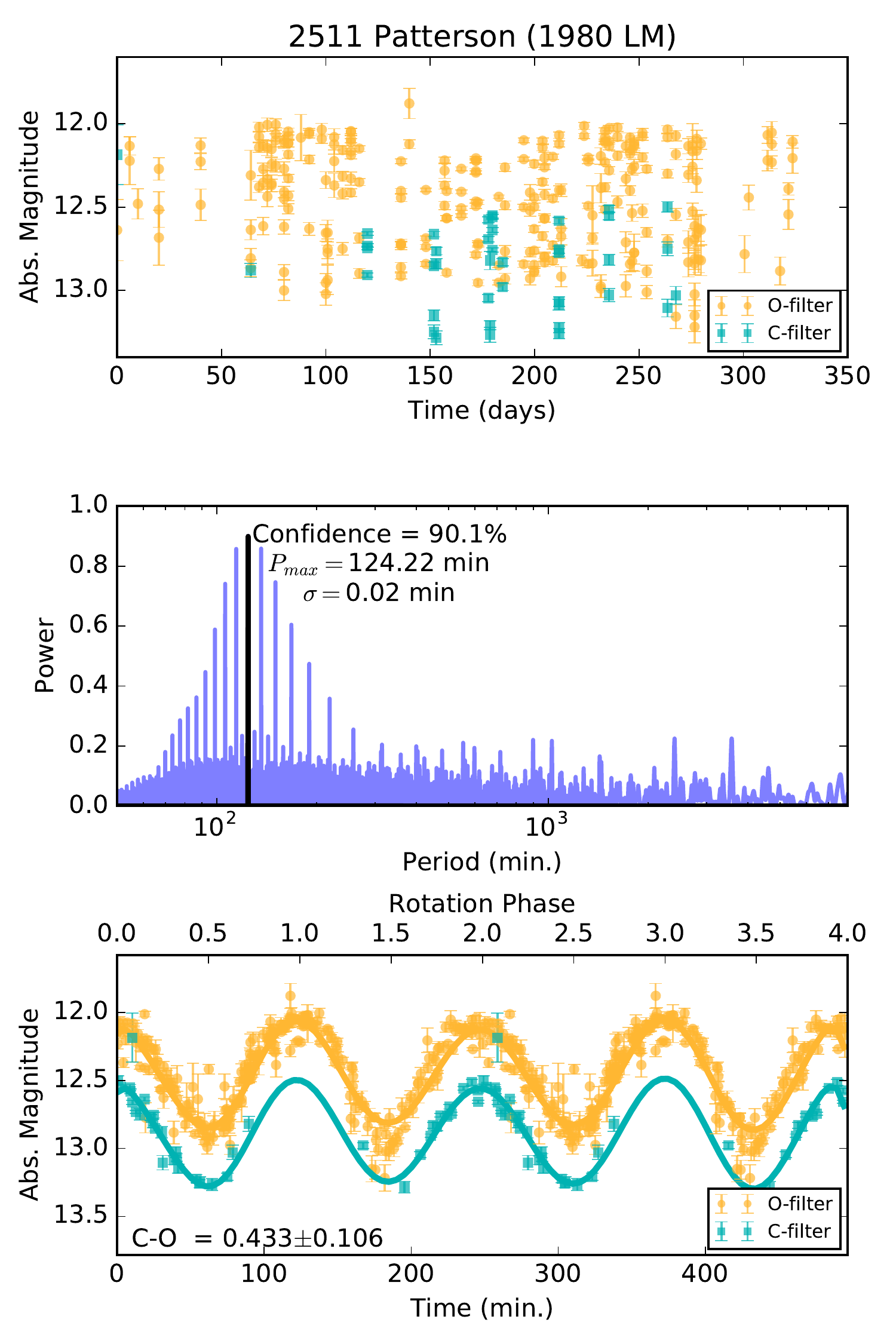}{0.3\textwidth}{(c)}
		}
		
		\caption{Example ATLAS photometric data (top plots), Lomb-Scargle periodograms of photometric data  (middle plots), and phased photometric data (bottom plots) for (a) Flora family member 1785 Wurm (1941 CD), (b) Koronis family member 1809 Prometheus (2522 P-L), and (c) Vesta family member 2511 Patterson (1980 LM). For the periodograms we also indicate the extracted light-curve period ($P_{max}=\frac{Rot. Period}{2}$), uncertainty in extracted period ($\sigma$), and confidence in extracted period. Fitted to the phased data are spline fits in the form of a high-order polynomial which are used to determine the color of each target (see Section \ref{sec:rot_color} for detail). The determined \textit{c-o} color value and the uncertainty is displayed in the bottom-left of each plot.} 
		\label{example_data}
	\end{center}
\end{figure*}

Rotation periods were extracted from the ATLAS data by generating a Lomb-Scargle periodogram \citep{Lomb1976,Scargle1982} of each target's \textit{o}-filter photometric data since this is the most abundant data out of the two filters used by ATLAS. We only considered ATLAS targets that had at least 30 photometric data points in the \textit{o}-filter and 10 photometric data points in the \textit{c}-filter. See top plots of Figure \ref{example_data} for example \textit{o}- and \textit{c}-filter photometric data of Flora family member 1785 Wurm (1941 CD), Koronis family member 1809 Prometheus (2522 P-L), and Vesta family member 2511 Patterson (1980 LM). Targets that had periodograms containing peaks with a confidence \citep{Zechmeister2009} larger than 50\% were flagged to have potentially extractable rotation periods. See middle plots of Figure \ref{example_data} for example periodograms which resulted in high confidence peaks.  The final extracted light-curve period is determined from the strongest periodogram peak and the uncertainty of this periodicity is determined by fitting a Gaussian function to the periodogram peak and using the RMS width ($\sigma$) as the uncertainty (see superimposed black curve and $\sigma$ of the fitted function in the periodogram plots in Figure \ref{example_data}). It has to be noted that because of the ATLAS observing cadence there is significant alias ambiguity in our periodograms due to the daily interval between observations. Therefore there is the likelihood that some of the periods we extract are offset from the actual period by a frequency that is a multiple of a day (24 hours). This is evident by the presence of multiple peaks in some of our periodograms that have a similar confidence. The rotation periods (twice the extracted light-curve period), uncertainty, and confidence in the periods are recorded in Table \ref{table_example_data}. 

To determine the colors of each ATLAS targets we fold both the \textit{c}- and \textit{o}-filter data with the extracted rotation period. A spline fit in the form of a high-order polynomial is fitted to the more-abundant \textit{o}-filter data with the fit weighted on the uncertainty of the photometric values. The same polynomial that is fitted to the \textit{o}-filter data, with the inclusion of a magnitude-offset as a variable fit parameter, is fitted to the  \textit{c}-filter data. The \textit{c-o} color is assigned as this fitted magnitude offset.  The uncertainty in this color value is derived from the  weighted (by uncertainty of the photometric values) standard deviation of the residuals of the two filter fits and the respective filter data. See bottom plots of Figure \ref{example_data} for examples of the results of this procedure with the \textit{c-o} color values and uncertainty displayed in bottom-left corner of each plot. This derived color value is not dramatically affected if by chance the folding is performed using an alias of the rotation period instead of the actual rotation period.

In a final step we remove photometric outliers and low-quality data from our data set by discarding targets that have a \textit{c-o} color that falls more than $3\sigma$ from the median measured \textit{c-o} color of the relevant family, and targets that had an uncertainty in measured  \textit{c-o} color larger than $130\%$ of the median uncertainty of the targets in the relevant family. The result was that we discarded on average  roughly $25\%$ of our targets for each family. The calculated \textit{c-o} colors and the uncertainties are recorded in Table \ref{table_example_data}.
 
\section{Taxonomic Determination}
\label{sec:tax_det}

\begin{figure*}[ht]
	\begin{center}
		\gridline{ 
			\fig{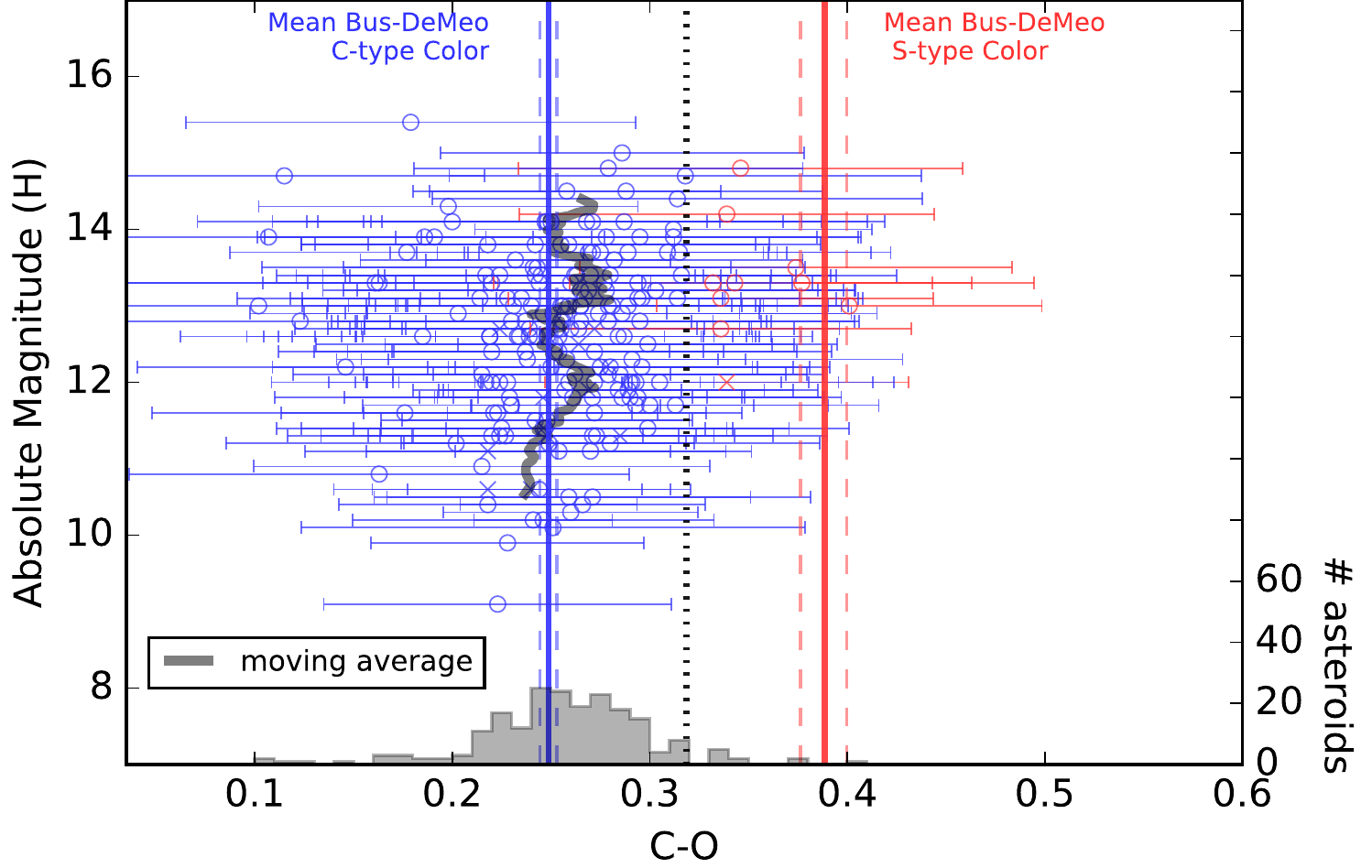}{0.45\textwidth}{(a) Themis}
			\fig{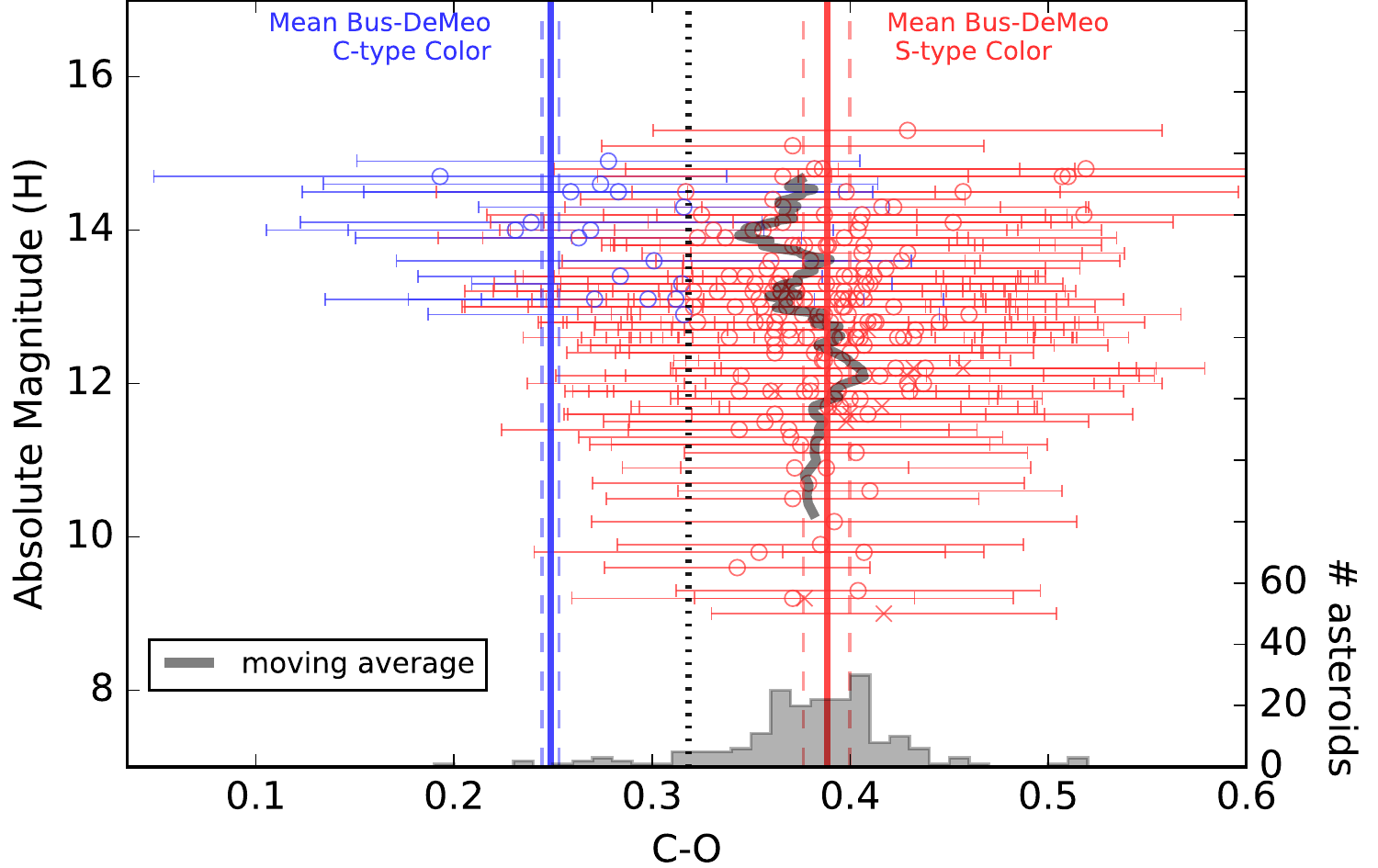}{0.45\textwidth}{(b) Koronis}
		}
		\caption{Plotted are the \textit{c-o} colors that could be extracted (see Section \ref{sec:rot_color}) from the ATLAS data set  of (a) Themis family members and (b) Koronis family members. The data points are colored depending on the final assigned taxonomy (see Section \ref{sec:tax_det}, S-type = red and C-type = blue). The Themis and Koronis families are two known C- and S-type Bus-DeMeo taxonomic families, respectively, and the color-histograms centered at the relevant mean Bus-DeMeo color therefore illustrates that the ATLAS \textit{c-o} color is a suitable parameter to distinguish between the C- and S-type taxonomies.}
		\label{themis_and_koronis}
	\end{center}
\end{figure*}

\begin{figure*}[ht]
	\begin{center}
		\gridline{ 
			\fig{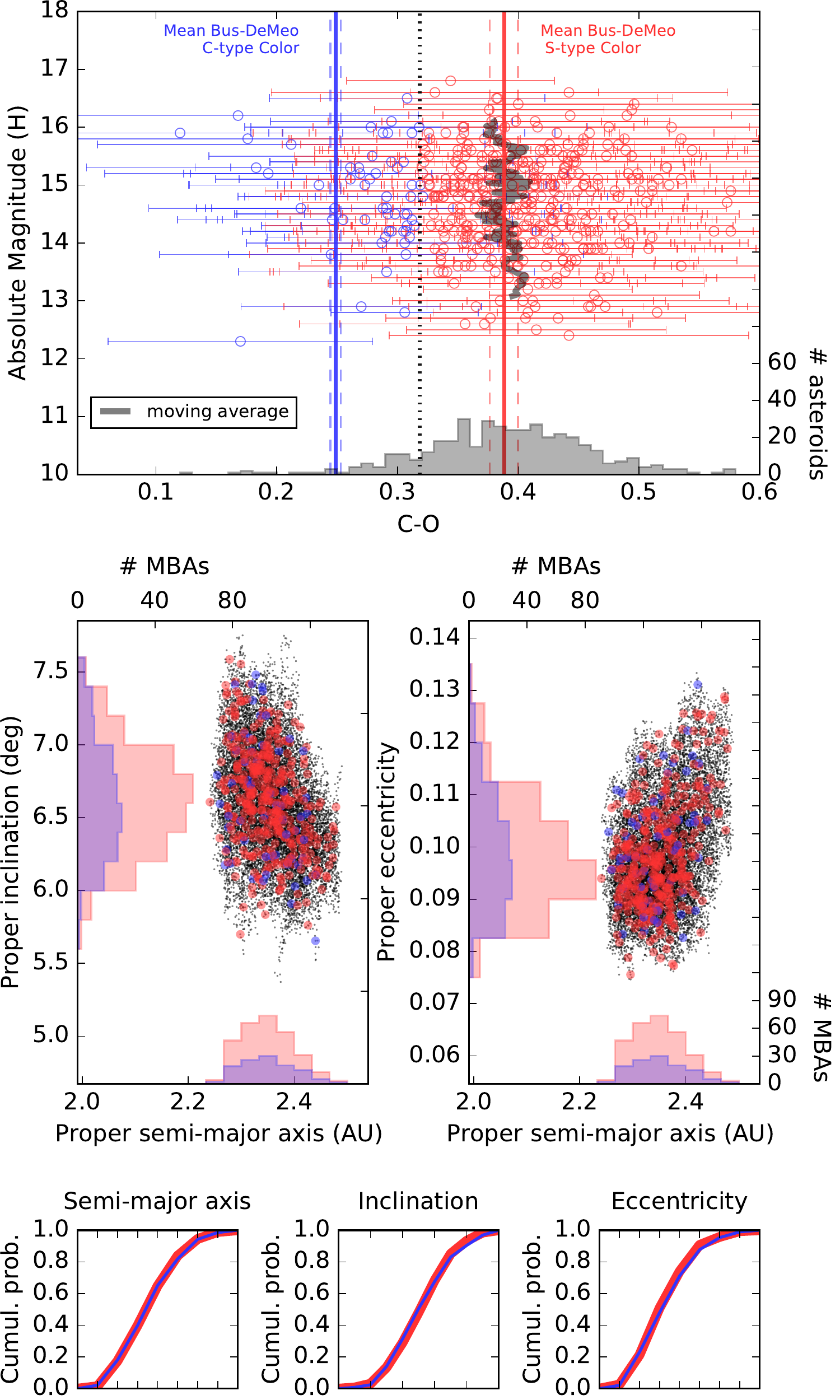}{0.45\textwidth}{(a) Vesta}
			\fig{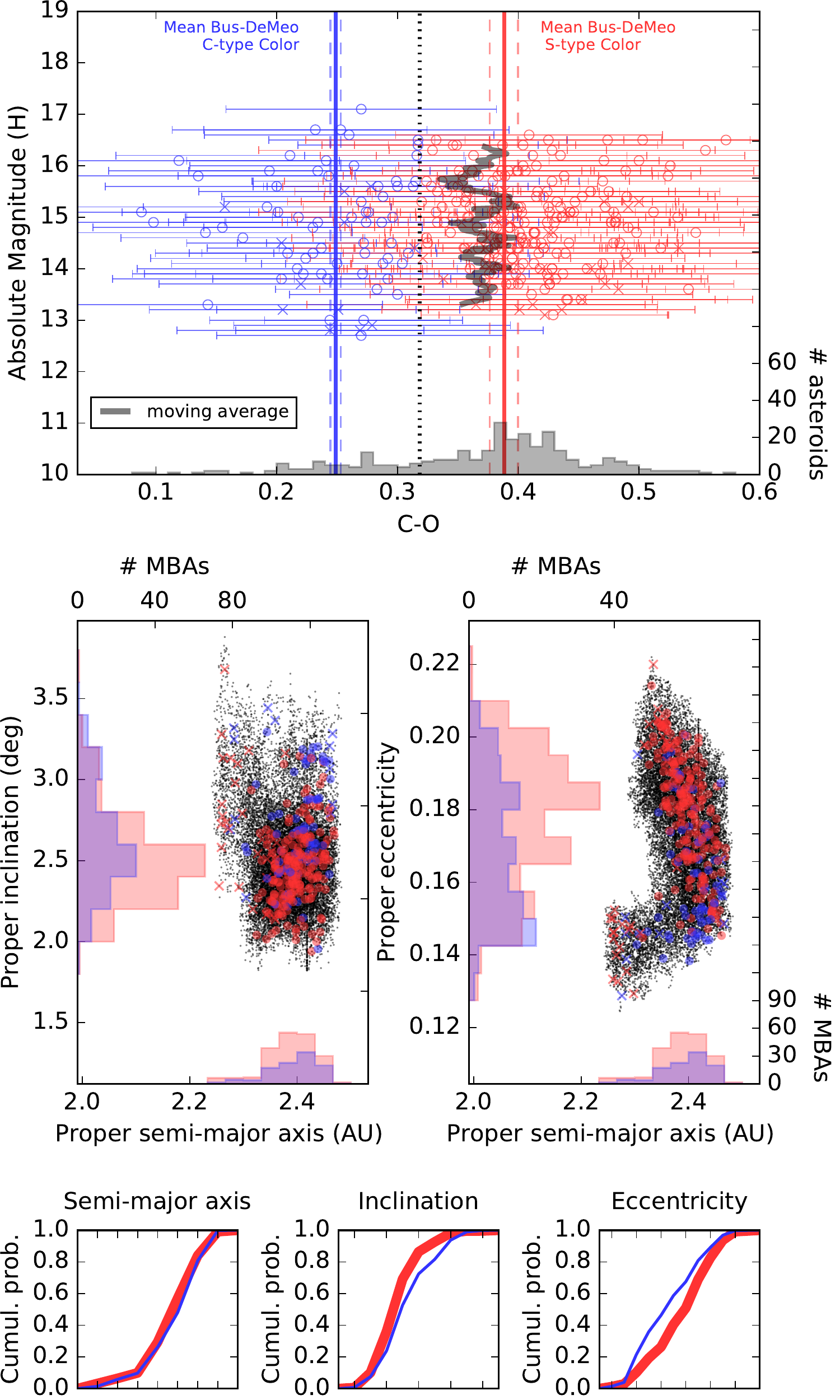}{0.45\textwidth}{(b) Nysa-Polana}
		}
		
		\caption{(Top) Plotted are the \textit{c-o} colors that could be extracted (see Section \ref{sec:rot_color}) from the ATLAS data set  of (a) Vesta family members and (b) Nysa-Polana family members. The data points are colored depending on the final assigned taxonomy (see Section \ref{sec:tax_det}, S-type = red and C-type = blue). (Middle)  Proper orbital elements of all ATLAS targets with all known family members from \cite{Nesvorny2015b}  plotted as small black data points in the background (suspected interlopers plotted with a $\times$-symbol). Also included are the histograms of the taxonomic probabilities (see Section \ref{sec:tax_det}) indicating the taxonomic dependence on orbital parameters. (Bottom)  The cumulative probabilities of the two taxonomies with respect to the three orbital parameters: broad red-line = S-type and narrow blue-line = C-type (see Section \ref{sec:vesta_nysa}). 
		} 
		\label{vesta_and_nysa-polana}
	\end{center}
\end{figure*}

For this study we limit our target classification to the two most prominent Bus-DeMeo taxonomic classes namely the silicate-rich S-type and the carbonaceous C-type which make up roughly 50\% and 35\% of the main-belt population respectively \citep{Erasmus2018,DeMeo2014,Bus2002}.  Because the Bus-DeMeo S- and V-type spectra  are difficult to distinguish using only two broadband filters in the visible region we consider those two both as S-type in this study. To decide which of these two classifications is the most likely for each of our targets we calculate the expected \textit{c-o} ATLAS color for the mean Bus-DeMeo  S- and C-type spectra provided by \cite{DeMeo2009}, and compare that to the target's measured \textit{c-o} color. The expected \textit{c-o} color is determined by convolving the ATLAS filter responses with the mean Bus-DeMeo S- and C-type spectra (see Figure 1 for filter responses and spectra) and use the convolution of the ATLAS filter responses with the Sun's spectrum as the zero-point magnitudes. The same was done for the the upper and lower bounds of the spectra to determine the uncertainty in the expected color. The expected \textit{c-o} color for a  S-type asteroid is $0.388^{+0.011}_{-0.012}$ and for a C-type $0.249^{+0.004}_{-0.004}$ (see solid blue and red vertical lines in Figures \ref{themis_and_koronis}, \ref{vesta_and_nysa-polana} and \ref{flora} with the upper and lower bounds indicated in dashed lines). 

\startlongtable
\begin{deluxetable*}{|c|l|c|r|r|rc|cc|}
	\tablecaption{ATLAS Colors, Extracted Rotation Periods, and Taxanomic Probabilities\tablenotemark{b} \label{table_example_data}}
	\tabletypesize{\tiny}		
	%\tablewidth{0pt}
	\tablehead{
		\colhead{No.} & \colhead{Target Name} & \colhead{Family} & \colhead{\textit{H}\tablenotemark{a} } & \colhead{\textit{c-o}} & \colhead{Rotation Period}& \colhead{Confidence} & \colhead{C-type} & \colhead{S-type}  \\
		\colhead{} & \colhead{} & \colhead{} & \colhead{(mag)} & \colhead{(mag)} & \colhead{(hours)}& \colhead{$\%$} & \multicolumn{2}{c}{(prob. in $\%$)} 
	}
	\startdata
	\input{table_for_tex_examples.txt}
	\enddata
	\tablenotetext{a}{H magnitude was obtained from \url{https://ssd.jpl.nasa.gov/horizons.cgi}}
	\tablenotetext{b}{This is only example data, the full table containing all 1612 targets is located in the Appendix}
	
\end{deluxetable*}

To assign the S- or C-type taxonomy to each of our targets we use a probabilistic approach using a Monte-Carlo method where we generate 10000 pseudo-measured colors with a normal distribution centered on the target's measured color and the width of the distribution equal to the uncertainty in the measured color (see \cite{Navarro_Meza_2019} for a similar approach using \textit{r}-\textit{i} colors to classify near-Earth asteroids). Each pseudo-measured color is classified as either S- or C-type depending on where it lies relative to the decision-line located at the color value midway between the two expected \textit{c-o} color values for an  S- and C-type asteroid (i.e., \textit{c-o}$ =0.319$, see dotted black vertical line in Figures \ref{themis_and_koronis}, \ref{vesta_and_nysa-polana} and \ref{flora}). The  pseudo-measured classifications are tallied to give the S- and C-type taxonomic probabilities for each target. The taxonomy with the highest probability is assigned to the target (data points in Figures \ref{themis_and_koronis}, \ref{vesta_and_nysa-polana} and \ref{flora} are color-coded depending on the final assigned taxonomy, S-type = red and C-type = blue).  We use the probabilities instead of the discrete classification for all our analyses in Section \ref{sec:vesta_nysa} and \ref{sec:flora}. The taxonomic probabilities for each target are recorded in Table \ref{table_example_data}.

\section{Validation of rotation periods and taxonomic determination}
\label{sec:validation}

To validate our procedures of extracting rotation periods and determining taxonomic types, we cross-reference our results with The Asteroid Light Curve Database \citep[LCDB;][Updated 2019 January 31]{Warner2009}\footnote{\url{http://www.MinorPlanet.info/lightcurvedatabase.html}} and find that $\sim10\%$  of our targets are also present in the LCDB. For the 262 targets present in both data sets we derive the same taxonomy as that reported in the LCDB for $85\%$ of the targets (for this we consider V-type and S-type as a match as well as B-type and C-type as a match) and also match the reported rotation periods within an error margin of $1\%$ for $76\%$ of our targets and within an error margin of $10\%$ for $89\%$ of our targets (for this comparison we also included the two alias periods adjacent to the main extracted period). 

As a second validation test, the Themis and Koronis families which are two well-known relatively pure C- and S-type Bus-DeMeo taxonomic families are used as benchmarks for our taxonomic determination. In  Figure \ref{themis_and_koronis} we plot the extracted \textit{c-o} colors of the Themis and Koronis family members within the ATLAS data set. The  data points are color coded depending on the final assigned taxonomy, S-type = red and C-type = blue. Of the 204 Themis targets we classify $95\%$ as C-type and the color histogram shown in Figure \ref{themis_and_koronis} (a) shows a clear peak centered at the expected mean Bus-DeMeo C-type color. Of the 197 Koronis targets we classify $91\%$ as S-type and the color histogram shown in Figure \ref{themis_and_koronis} (b) shows a clear peak centered at the expected mean Bus-DeMeo S-type color. These two plots demonstrate that the ATLAS \textit{c-o} colors that we extract are a sufficient parameter to distinguish between the C- and S-type taxonomies. 

\section{Mixture of Taxonomies in the Vesta and Nysa-Polana Families}
\label{sec:vesta_nysa}

The Vesta family, a V-type family, has a spectral signature similar to the spectral shape of S-type asteroids but with the distinct S-type absorption feature at $1\mu$m enhanced in V-types.  The Nysa-Polana family is also a known S-type family. However, observations have indicated that both contain a significant number of family members with  spectra or color different to the majority of family members \citep{Erasmus2019}. This could be ascribed to the parent body of the Vesta family originally consisting of a differentiated object \citep{Russell2012}  and hence a plausible reason behind the range in colors observed within the Vesta family. Through albedo studies the Nysa-Polana family has been identified to consist of two subgroups overlapping in proper orbital space \citep{Cellino2001} i.e., two nested families of which one shows S-type spectral characteristics and the other B-type spectral characteristics. The Bus-DeMeo B-type spectrum is similar to the spectral shape of C-type asteroids, flat and featureless, but with a slightly bluer slope in the visible. The top set of plots in Figure \ref{vesta_and_nysa-polana} are the extracted \textit{c-o} colors of the Vesta and Nysa-Polana family members within the ATLAS data set. The two-taxonomic mix is clearly evident in both families with only $\sim70\%$ of Vesta targets having \textit{c-o} colors in the vicinity of the expected S(V)-type color while the Nysa-Polana targets have a $1:3$ split between \textit{c-o} colors  that match the expected C(B)-type and S-type colors respectively. 

By plotting the distributions of the two taxonomies as a function of proper orbital space (see middle set of plots in Figure \ref{vesta_and_nysa-polana}), the differing causes behind the taxonomic diversity seen in both of these families can be distinguished from one another. The assigned taxonomy of Vesta family members have no dependence on proper orbital space (consistent with a differentiated parent body origin) while the Nysa-Polana ``B-type subgroup" is positioned at a lower eccentricity (and to a lesser extend higher inclination) than the Nysa-Polana ``S-type subgroup" (consistent with two nested families with slightly differing orbital parameters). This taxonomic dependence (or independence) on orbital parameters is highlighted by performing Kolmogorov-Smirnov (K-S) tests and plotting the resultant cumulative probabilities of the two taxonomies with respect to the three orbital parameters (see bottom plots of Figure \ref{vesta_and_nysa-polana}, broad red-line = S-type and narrow blue-line = C-type). The two taxonomies in the Vesta family have identical cumulative probabilities for the three orbital parameters (K-S statistics for semi-major axis, inclination, and  eccentricity  of 0.08, 0.04, and 0.01) while the two taxonomies in the Nysa-Polana family have obvious differing cumulative probabilities in eccentricity  (K-S statistics for semi-major axis, inclination, and  eccentricity  of 0.05, 0.12, and 0.15). We use the cumulative probabilities of these two families as benchmarks for ascertaining the likely cause of the taxonomic diversity also seen in the Flora family.

\section{Taxonomic Diversity of the Flora Family}
\label{sec:flora}

\begin{figure}[ht]
	\begin{center}
		\includegraphics[width=0.45\textwidth]{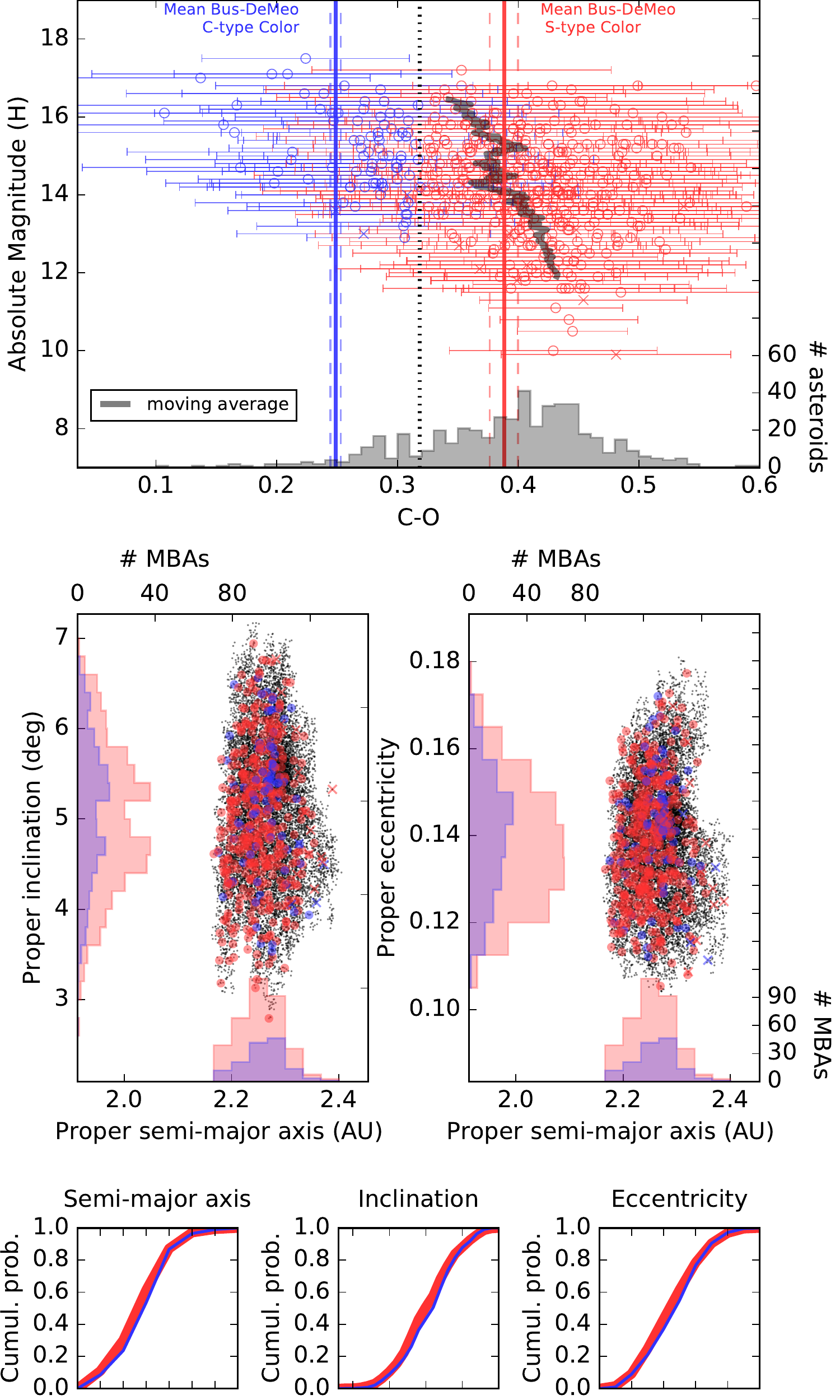}
		\caption{(Top) Plotted are the \textit{c-o} colors that could be extracted (see Section \ref{sec:rot_color}) from the ATLAS data set  of Flora family members. The data points are colored depending on the final assigned taxonomy (see Section \ref{sec:tax_det}, S-type = red and C-type = blue). (Middle)  Proper orbital elements of all ATLAS targets with all known family members from \cite{Nesvorny2015b}  plotted as small black data points in the background (suspected interlopers plotted with a $\times$-symbol). Also included are the histograms of the taxonomic probabilities (see Section \ref{sec:tax_det}) indicating the taxonomic dependence on orbital parameters. (Bottom)  The cumulative probabilities of the two taxonomies with respect to the three orbital parameters: broad red-line = S-type and narrow blue-line = C-type (see Section \ref{sec:vesta_nysa})}
		\label{flora}
	\end{center}
\end{figure}

 The top plot in Figure \ref{flora} shows the the extracted \textit{c-o} colors of the Flora family members within the ATLAS data set. As is the case with the Vesta and Nysa-Polana families, the Flora family also displays a significant taxonomic mix, or at least a significant range of colors. Roughly $70\%$ of the targets have \textit{c-o} colors consistent with the expected S-type color with the remaining targets displaying bluer C-like colors. This is a similar proportion to that of the Vesta family. The cumulative plots of the two colors in the Flora family (see bottom of Figure \ref{flora}) are identical for the three orbital parameters (as is the case for the Vesta family) as opposed to indicating a nested family as is the case for the Nysa-Polana family. Therefore a nested family is unlikely as the source of the taxonomic mix in the Flora family.

\section{Discussion}

There are several possible interpretations for the taxonomic mix observed for the Flora family. Plotting the moving average of the colors as a function of H-magnitude (size) reveals a strong color dependence on size for the Flora family targets (see Figure \ref{flora}). Since the smaller bodies are probably younger (more recently created through collisions) and therefore have fresher surfaces, one clear possibility is that the older surfaces on the larger bodies have been reddened through space weathering, whereas the younger surfaces on smaller bodies are unweathered and therefore less red. \cite{Thomas2012,Thomas2011a} report that small ($H\approx15$) Koronis family objects are bluer than large ($H\approx12$) Koronis family objects and attribute this difference to space weathering. We also observe a hint of this effect at a similar H-magnitude range in our Koronis family observations (see Figure \ref{themis_and_koronis} (b)). However this effect is subtle and can not explain the large color range we observe for the Flora family (compare color ranges as a function of H-magnitude for the Koronis and Flora  families in Figure \ref{themis_and_koronis} (b) and \ref{flora})

Another possibility is that, like the Vesta family, the heterogeneous Flora family may show evidence of a differentiated parent body. This is in agreement with a differentiated-body speculation by \cite{Gaffey1984} based on several mineralogic and petrologic parameters derived for the surface material of asteroid Flora. Spectral data from the Galileo spacecraft on its flyby of  Flora-family member 951 Gaspra  also showed an olivine/pyroxene ratio that points to Gaspra being a fragment of at least a partially differentiated parent body \citep{Veverka1994}. However,  various work \citep{Vernazza2008,Dunn2013,Vernazza2014} has shown that Flora family is likely the source of the undifferentiated LL chondrites, making this interpretation potentially problematic.

Alternatively, the C-like asteroids in the Flora family may simply be interlopers in the family membership list, though this would imply a relatively high contamination fraction of 28.3\% which is significantly higher than the estimation for our data set of only 4.5\% using the model of \cite{Nesvorny2015b}.

Finally, two additional possible causes for the C-like colors that are observed could be due to shocked material and impact melt, which \cite{Kohout2014} and \cite{Reddy2014} have shown can make LL chondrite (S-like) spectra appear more C-like, or that a large member of the original S-type Flora family was simply impacted by a C-type asteroid that shattered, producing an embedded family within the Flora family that has a separate parent body (here we define an embedded family as one that was created through a collision between a family member and a non-family member whereas a nested family occupies similar orbital element space but has an independent origin). Neither of these interpretations can be excluded by the taxonomic independence on orbital parameters we observe. The strong color dependence on size of our observed Flora family targets can be explained if the C-type impactor was smaller ($H \gtrapprox 12$) than the original S-type Flora parent body.

These topics are discussed in further detail in a forthcoming paper (Sun et al., in prep.).
\section{Conclusion}

We have reported the \textit{c-o} colors (and rotation periods) for identified Flora, Vesta, Nysa-Polana, Themis, and Koronis family members within the historic data set (2015-2018) of ATLAS. By using a probabilistic approach we also classify our targets as either S-type or C-type and compare to previously reported taxonomies and rotation periods to validate our methodologies. We use the Themis (a C-type family) and Koronis (a S-type family) targets in the ATLAS data set to demonstrate that the ATLAS \textit{c-o} color is a sufficient parameter to distinguish between the C- and S-type taxonomies. 

We find that the Flora family has a significant mixture of red (S-like) and blue (C-like) colors and a compelling observed H-magnitude dependence on the colors and we propose several possible causes of this range in colors.

\section*{Acknowledgments}

This work has made use of data from the Asteroid Terrestrial-impact Last Alert System (ATLAS) project. ATLAS is primarily funded to search for near earth asteroids through NASA grants NN12AR55G, 80NSSC18K0284, and 80NSSC18K1575; byproducts of the NEO search include images and catalogs from the survey area.  The ATLAS science products have been made possible through the contributions of the University of Hawaii Institute for Astronomy, the Queen's University Belfast, the Space Telescope Science Institute, and the South African Astronomical Observatory.

This work is partially supported by the South African National Research Foundation (NRF), the Arizona Board of Regents' Regents Innovation Fund, the National Aeronautics and Space Administration (NASA), and by a grant from NASA's Office of the Chief Technologist. 

\bibliography{N_Erasmus}

\appendix 
\vspace*{\fill}
\section{All Data}
\vspace*{\fill}

\startlongtable
\begin{deluxetable*}{|c|l|c|r|r|rc|cc|}
	\tablecaption{ATLAS Colors, Extracted Rotation Periods, and Taxanomic Probabilities \label{table_all_data}}
	\tabletypesize{\tiny}		
	%\tablewidth{0pt}
	\tablehead{
		\colhead{No.} & \colhead{Target Name} & \colhead{Family} & \colhead{\textit{H}\tablenotemark{a} } & \colhead{\textit{c-o}} & \colhead{Rotation Period}& \colhead{Confidence} & \colhead{C-type} & \colhead{S-type}  \\
		\colhead{} & \colhead{} & \colhead{} & \colhead{(mag)} & \colhead{(mag)} & \colhead{(hours)}& \colhead{$\%$} & \multicolumn{2}{c}{(prob. in $\%$)} 
	}
	\startdata
	\input{table_for_tex.txt}
\enddata
	\tablenotetext{a}{H magnitude was obtained from \url{https://ssd.jpl.nasa.gov/horizons.cgi}}
\end{deluxetable*}

\end{document}

%% file: table_for_tex_examples.txt
0002&167 Urda&Koronis&9.2&0.38$\pm$0.06&13.060$\pm$0.009&88&14&86\\
0008&277 Elvira&Koronis&9.8&0.41$\pm$0.04&29.676$\pm$0.046&86&1&99\\
0018&540 Rosamunde (A904 PE)&Flora&10.8&0.44$\pm$0.06&9.348$\pm$0.004&88&1&99\\
0606&10541 Malesherbes (1991 YX)&Flora&14.3&0.29$\pm$0.06&20.224$\pm$0.051&94&71&29\\
0905&23266 (2000 YP50)&Nysa-Polana&14.7&0.41$\pm$0.04&11.114$\pm$0.020&96&1&99\\

%% file: table_for_tex.txt
0001&158 Koronis&Koronis&9.3&0.40$\pm$0.09&14.204$\pm$0.006&73&18&82\\
0002&167 Urda&Koronis&9.2&0.38$\pm$0.06&13.060$\pm$0.009&88&14&86\\
0003&208 Lacrimosa&Koronis&9.0&0.42$\pm$0.09&14.087$\pm$0.010&84&13&87\\
0004&222 Lucia&Themis&9.1&0.22$\pm$0.09&9.371$\pm$0.005&58&86&14\\
0005&243 Ida&Koronis&9.9&0.39$\pm$0.10&4.225$\pm$0.001&74&26&74\\
0006&254 Augusta&Flora&12.1&0.42$\pm$0.12&5.895$\pm$0.002&85&18&82\\
0007&263 Dresda&Koronis&10.2&0.39$\pm$0.12&12.450$\pm$0.055&72&28&72\\
0008&277 Elvira&Koronis&9.8&0.41$\pm$0.04&29.676$\pm$0.046&86&1&99\\